\documentstyle[fleqn]{article}
\let\up=\uparrow
\let\down=\downarrow
\let\t=\dagger
\let\la=\lambda
\let\eps=\varepsilon
\def\Spur{\mathop{\rm Sp}}
\def\sgn{\mathop{\rm sgn}}
\def\Tr{\mathop{\rm Tr}}

\begin{document}

\begin{titlepage}
\rightline{PDMI PREPRINT --- 5/1998}
\vspace*{.5in}
\begin{center}
{\Large\bf
Temperature correlators in the one-dimensional\\[6pt]
Hubbard model in the strong coupling limit
}\\[.5in]
{\large\sc
A.\,G.\,Izergin%
${}^{\star}$\footnote{E-mail: izergin@pdmi.ras.ru},
A.\,G.\,Pronko%
${}^{\star}$\footnote{E-mail: agp@pdmi.ras.ru},
N.\,I.\,Abarenkova%
${}^{\star\star}$\footnote{E-mail: aiv@iva.usr.pu.ru}}\\[.5in]
{\large\sc ${}^{\star}$}{\it Sankt Petersburg Department
of V.\,A.\,Steklov Mathematical Institute,\\
Fontanka 27, 191\,011 Sankt Petersburg, Russia}\\[6pt]
{\large\sc ${}^{\star\star}$}{\it Theoretical Physics Department,
Sankt Petersburg State University,\\
Ulyanovskaya 1, 198\,904 Sankt Petersburg, Russia}
\end{center}
\vspace{.5in}
\hrule\vspace{.3in}

\centerline{{\bf ABSTRACT}}
\bigskip

We consider the one-dimensional Hubbard model
with the infinitely strong repulsion.
The two-point dynamical temperature correlation functions
are calculated. They are
represented as Fredholm determinants of linear integrable integral
operators.

\vspace{.3in}\hrule
\end{titlepage}

\section{Introduction}

The one-dimensional Hubbard model \cite{H-63} is one of the
most interesting and
important model of strongly correlated electrons (see, e.g., book
\cite{KE-94}). The hamiltonian of the model with the chemical potential
$h$ in the external constant field $B$ describes interacting fermions
on the one-dimensional periodic lattice of length $L$:
\begin{eqnarray} \label{H_U}
\lefteqn{
H_U=-\sum_{x=1}^{L} \sum_{\alpha=\up,\down}^{}
\Bigl( c_{x,\alpha}^\t c_{x-1,\alpha}^{}
+c_{x-1,\alpha}^\t c_{x,\alpha}^{} \Bigr)
}
\nonumber\\ &&
-h\sum_{x=1}^{L}(n_{x,\up}+n_{x,\down})
+B\sum_{x=1}^{L}(n_{x,\up}-n_{x,\down})
+U\sum_{x=1}^{L} n_{x,\up} n_{x,\down}.
\end{eqnarray}
Here $c_{x,\alpha}^{}$ and $c_{x,\alpha}^\t$ ($x=1,\dots,L$;
$\alpha=\up,\down$; $c_{0,\alpha}^{}\equiv c_{L,\alpha}^{}$;
$c_{0,\alpha}^\t\equiv c_{L,\alpha}^\t$) are canonical
fermion fields with the anticommutation relations
\begin{equation} \label{anti}
[c_{x,\alpha}^{}, c_{y,\beta}^\t]_{+}
=\delta_{x,y}\delta_{\alpha,\beta}
\end{equation}
The density operators of electrons with spin up
($\alpha=\up$) and down ($\alpha=\down$) are
$n_{x,\alpha}^{}=c_{x,\alpha}^\t c_{x,\alpha}^{}$.
The solution of this model by means of the two-component (nested)
Bethe Anzatz \cite{Y-67,G-67} was given in \cite{LW-68}.

The temperature correlation functions are defined in a standard way as
temperature normalized mean values. In the present paper we consider
two-point correlation functions of canonical field operators,
\begin{eqnarray} \label{cc}
\lefteqn{
\langle c^\t_{x,\alpha}(t) c^{}_{0,\alpha}(0) \rangle^{(T,L)}
={\Spur\Bigl[ e^{- H_U /T}
c^\t_{x,\alpha}(t) c^{}_{0,\alpha}(0)\Bigr]\over
\Spur\Bigl[e^{- H_U /T}\Bigr]},
}
\nonumber\\
\lefteqn{
\langle c^{}_{x,\alpha}(t) c^\t_{0,\alpha}(0) \rangle^{(T,L)}
={\Spur\Bigl[ e^{- H_U /T}
c^{}_{x,\alpha}(t) c^\t_{0,\alpha}(0)\Bigr]\over
\Spur\Bigl[e^{- H_U /T}\Bigr]},
}
\end{eqnarray}
and of density operators,
\begin{equation} \label{nn}
\langle n_{x,\alpha}(t) n_{0,\beta}(0) \rangle^{(T,L)}
={\Spur\Bigl[ e^{- H_U /T}
n_{x,\alpha}(t) n_{0,\beta}(0)\Bigr]\over
\Spur\Bigl[e^{- H_U /T}\Bigr]},
\end{equation}
where $\alpha,\beta=\up,\down$.
The dependence on time is introduced as usual,
\begin{eqnarray} \label{time}
\lefteqn{
c^{}_{x,\alpha}(t) = e^{i H_U t} c^{}_{x,\alpha} e^{-i H_U t},\quad
c^\t_{x,\alpha}(t) = e^{i H_U t} c^\t_{x,\alpha} e^{-i H_U t},
}
\nonumber\\
\lefteqn{
n^{}_{x,\alpha}(t) = e^{i H_U t} n^{}_{x,\alpha} e^{-i H_U t}.
}
\end{eqnarray}
The trace $\Spur$ is taken in the Fock space ${\cal F}$ of dimension
$4^L$ where the hamiltonial of the Hubbard model acts.
The correlation functions are mostly interesting in the thermodynamic
limit ($L\to\infty$; $h$ and $B$ are kept fixed):
\begin{equation} \label{tdlim}
\langle \cdots \rangle^{(T)}
=\lim_{L\to\infty}
\langle \cdots \rangle^{(T,L)}.
\end{equation}

Our aim is to present the results on calculation of the
correlation functions in the Hubbard model in the strong coupling
limit, $U\to +\infty$.  The two-point temperature correlation
functions are
represented as Fredholm determinants of linear integral operators of a
special kind (the ``integrable'' integral operators, in the sence of the
paper \cite{IIKS-93}).

It is to be mentioned that
the recent progress in calculating correlation functions of quantum
solvable models is based on the fact that they are governed by
classical integrable differential equations. That the language of
classical differential equations is quite natural for the description
of quantum correlation functions was realized a time ago
\cite{Tr1,Tr2,Tr3,Jimbo}.  The idea of the approach suggested in
\cite{IIK-89,IIKS-90,IIKS-93} is to consider the Fredholm determinant
in the representation for a correlation function of a quantum
integrable model as a tau-function for a classical integrable system
(see also the book \cite{KBI} where the results for the simplest model
of one-dimensional impenetrable bosons are reviewed). The necessary
first step (which is also of interest by itself) in this approach is to
represent the correlation function as the Fredholm determinant of a
linear ``integrable'' integral operator. The first
determinant representation of this kind was given in \cite{L-64,L-66} for
the equal-time temperature correlators of the one-dimentional impenetrable
bosons. For the two-component one-dimensional impenetrable Bose and Fermi
gases the representations of this type were obtained recently
in papers \cite{IP-PLA,IP-NPB}.

The Hubbard model is a lattice analogue of the two-component
Fermi gas. It should be noted that the physics of the Hubbard model at
$U=+\infty$ has some interesting properties
(see, e.g., \cite{FK-91}). In particular, the ground state becomes
degenerate at zero external field, i.e., the point $B=0$ is the point
of the phase transition (the same take place in the Fermi gas as well).

We use the technique of calculating temperature correlation
functions of exactly solvable quantum two-component models
with infinitely strong coupling developed
in papers \cite{IP-PLA,IP-NPB} for one-dimensional
two-component Bose and Fermi gases. The starting point of the approach
is using the eigenfunctions of the $XX0$ spin chain
with the periodic boundary conditions to solve the auxiliarly
lattice problem of the nested Bethe Ansatz.
This gives an explicit expression for the eigenstates of the
two-component models at the point of infinite coupling.

\section{The hamiltonian and eigenstates at $U=\infty$}

In the strong coupling limit, the states with double occupancies
(i.e., the states with a pair of electrons at least
at one cite of the lattice) have infinite energy and therefore are
absent in the physical space of states.
The dynamics of the Hubbard model in
this limit is described by an effective hamiltonian which can be written
in the form \cite{M-91,S-91}
\begin{eqnarray} \label{Heff}
\lefteqn{
H={\cal P}\Biggl[
-\sum_{x=1}^{L} \sum_{\alpha=\up,\down}^{}
\Bigl( c_{x,\alpha}^\t c_{x-1,\alpha}^{}
+c_{x-1,\alpha}^\t c_{x,\alpha}^{} \Bigr)
}
\nonumber\\&&
-h\sum_{x=1}^{L}(n_{x,\up}+n_{x,\down})
+B\sum_{x=1}^{L}(n_{x,\up}-n_{x,\down})
\Biggr]{\cal P},
\end{eqnarray}
where
\begin{equation} 
{\cal P}=\prod_{x=1}^{L}(1-n_{x,\up}n_{x,\down}).
\end{equation}
The projector  ${\cal P}$ extracts the physical space of states
${\cal H}$ of dimention $3^L$ from the space
${\cal F}$ (of dimention $4^L$) in which the canonical Hubbard operators
act.

Eigenstates of the effective hamiltonian
$H$ (all belonging to the space ${\cal H}$)
can be constructed analogously to the case of the two-component
impenetrable Fermi gas \cite{IP-PLA,IP-NPB}. The eigenstates
have the form \cite{LW-68}
\begin{eqnarray} \label{PsiNM}
\lefteqn{
|\Psi_{N,M}(k;\la)\rangle =
\sum_{z_1,\dots,z_N=1}^{L}
\sum_{\alpha_1,\dots,\alpha_N=\up,\down}^{}
\chi_{N,M}^{\alpha_1\dots\alpha_N} (z_1,\dots,z_N|k;\la)
}
\nonumber\\ &&
c_{z_1,\alpha_1}^\t \dots c_{z_N,\alpha_N}^\t |0\rangle,
\end{eqnarray}
where the Fock vacuum $|0\rangle$ is defined as usual,
$c_{z,\alpha}^{}|0\rangle=0$, $\langle 0|c_{z,\alpha}^\t=0$,
$\langle 0|0 \rangle=1$. The eigenstates are parametrized by
set of $N$ unequal real numbers (quasimomenta of $N$ particles),
$k\equiv k_1,\dots,k_N$, and the set of $M$ unequal real
numbers (quasimomenta of the auxiliarly lattice problem),
$\la\equiv\la_1,\dots,\la_M$.
The number $N$ is the total number of electrons in the state while
the number $M$ is the number of electrons with spin down.
Thus $M$ of $N$ values $\alpha$ in (\ref{PsiNM}) are $\down$,
and $N-M$ are $\up$.
The momenta $k$ and momenta $\la$  are not arbitrary but
must obey the system of the Bethe equations (the nested Bethe Ansatz)
\cite{Y-67,G-67,LW-68}.
In the case $U=+\infty$ using the $XX0$ basis for the auxiliarly problem
results in the simplified form of the Bethe equations,
\begin{eqnarray} \label{Bethe}
\lefteqn{
e^{ik_a L}=e^{i\Lambda},\qquad  a=1,\dots,N,
}
\nonumber\\
\lefteqn{
e^{i\la_b N}= (-1)^{M+1},\qquad b=1,\dots,M,
}
\end{eqnarray}
where
\begin{equation} 
\Lambda\equiv \sum_{b=1}^{M} \la_b.
\end{equation}
This system can be solved explicitly; the equations are separated by the
substitution
\begin{equation} 
k_a = \tilde k_a +{\Lambda \over L}.
\end{equation}
The permitted values of the momenta are
\begin{eqnarray} 
\lefteqn{
(\tilde k_a)_j
={2\pi \over L} j,\qquad j=0,\dots,L-1,
}
\nonumber\\
\lefteqn{
(\la_b)_l
={2\pi \over N}\left(-{N \over 2}+{1+(-1)^{N-M} \over 4}+l\right),
\qquad  l=0,\ldots,N-1.
}
\end{eqnarray}
The wave function $\chi_{N,M}$ is
\begin{eqnarray} 
\lefteqn{
\chi_{N,M}^{\alpha_1\dots\alpha_N}(z_1,\dots,z_N|k;\la)
}
\nonumber\\&&
= {1 \over N!}
\left[\sum_{P}^{}
\xi^{\alpha_{P_1}\dots\alpha_{P_N}}_{N,M}(\la)
\theta(z_{P_1}<\dots<z_{P_N}) \right]
{\det}_N \{e^{ik_a z_b}\}.
\end{eqnarray}
The sum here is taken over the permutations of
$N$ numbers, $P:(1,\dots,N)\to(P_1,\dots,P_N)$. The function
$\theta(z_1<\dots<z_N)$
is equal to $1$ if $z_1<\dots<z_N$ and is equal to zero otherwise.
The determinant ${\det}_N \{e^{ik_a z_b}\}$ denotes the determinant
of the $N\times N$ matrix with elements $e^{ik_a z_b}$.
The spin part of the wave function is described by components of the
$2^N$-dimensional vector $|\xi_{N,M}(\la))$. As in the case of the
two-component impenetrable gas, we use the vectors $|\xi_{N,M}(\la))$
to be eigenvectors of the hamiltonian of the $XX0$ spin chain in the
form given in \cite{CIKT-93}:
\begin{equation} 
|\xi_{N,M}(\la))
=\sum_{n_1,\dots,n_M=1}^{N} \varphi_{N,M}(n_1,\dots,n_M|\la)\
\sigma^{-}_{(n_1)}\dots \sigma^{-}_{(n_M)}|\!\Uparrow_N),
\end{equation}
with the wave function
\begin{equation} 
\varphi_{N,M}(n_1,\dots,n_M|\la)
={1 \over M!}
\left[\prod_{1\le j<l \le N}^{} {\sgn}(n_l-n_j)\right]\
{\det}_M \{e^{i\lambda_a n_b}\}.
\end{equation}
Pauli matrices are defined as usual,
\begin{equation} 
\sigma^{+}=\pmatrix{ 0 & 1 \cr 0 & 0},\qquad
\sigma^{-}=\pmatrix{ 0 & 0 \cr 1 & 0}.
\end{equation}
The vacuum $|\!\Uparrow_N)$ of the spin chain with  $N$ sites
is defined as $\sigma^{+}_{(n)} |\!\Uparrow_N)=0$,
$(\Uparrow_N\!|\sigma^{-}_{(n)}=0$, $(\Uparrow_N\!|\!\Uparrow_N)=1$,
i.e.,
\begin{equation} 
|\!\Uparrow_N)=\otimes_{n=1}^{N} \pmatrix{ 1\cr 0}_{(n)}.
\end{equation}
Eigenvalues of the hamiltonian (\ref{Heff}) on the eigenstates
(\ref{PsiNM}),
\begin{equation} 
H\ |\Psi_{N,M}(k,\la)\rangle
=E_{N,M}(k)\ |\Psi_{N,M}(k,\la)\rangle,
\end{equation}
are
\begin{equation} 
E_{N,M}(k)=\sum_{a=1}^{N}\eps(k_a)-hN+B(N-2M),
\end{equation}
where $\eps(k)$ is one-particle dispersion
\begin{equation} \label{eeps}
\eps(k_a)=-2\cos k_a= -2\cos\left(\tilde k_a +{\Lambda\over L}\right).
\end{equation}
Thus, the eigenenergies depend on $\la_1,\dots,\la_M$ via their sum,
$\Lambda$. It is used essensially in our approach.

The states $|\Psi_{N,M}(k;\la)\rangle$ form a complete orthogonal set
in the physical space ${\cal H}$ (of dimention $3^L$); the normalization
is
\begin{equation} 
\langle \Psi_{N,M}(k;\la)|\Psi_{N,M}(k;\la) \rangle =L^N N^M.
\end{equation}

The effect of the infinitely strong interaction consists in contraction
of observables (quantum operators) onto the space ${\cal H}$.
The unit operator in ${\cal H}$ is just the projector ${\cal P}$ in
${\cal F}$. Due to the completness, this operator possesses
the decomposition
\begin{equation} \label{P}
{\cal P}=\sum_{\rm states}^{}
|\Psi_{N,M}(k;\la)\rangle\frac{1}{L^N N^M}\langle\Psi_{N,M}(k;\la)|,
\end{equation}
where the summation is performed over all $3^L$ basis states in ${\cal H}$,
\begin{equation} 
\sum_{\rm states}^{}\equiv \sum_{N=0}^{L} \sum_{M=0}^{N}
\sum_{k_1<\dots<k_N \atop \la_1<\dots<\la_M}.
\end{equation}
Due to (\ref{P}), the effective hamiltonian $H$ (\ref{Heff})
can be also defined by its spectral decomposition
\begin{equation} 
H=\sum_{\rm states}^{}
|\Psi_{N,M}(k;\la)\rangle
\frac{E_{N,M}(k)}{L^N N^M}\langle\Psi_{N,M}(k;\la)|.
\end{equation}

The temperature correlation functions in the strong coupling limit acquire
the form:
\begin{eqnarray} \label{ccnn}
\lefteqn{
\langle c^\t_{x,\alpha}(t) c^{}_{0,\alpha}(0) \rangle^{(T,L)}
={\Tr\Bigl[ e^{- H/T}
\tilde c^\t_{x,\alpha}(t) \tilde c^{}_{0,\alpha}(0)\Bigr]\over
\Tr\Bigl[e^{- H/T}\Bigr]},
}
\nonumber\\
\lefteqn{
\langle c^{}_{x,\alpha}(t) c^\t_{0,\alpha}(0) \rangle^{(T,L)}
={\Tr\Bigl[ e^{- H/T}
\tilde c^{}_{x,\alpha}(t) \tilde c^\t_{0,\alpha}(0)\Bigr]\over
\Tr\Bigl[e^{- H/T}\Bigr]},
}
\nonumber\\
\lefteqn{
\langle n_{x,\alpha}(t) n_{0,\beta}(0) \rangle^{(T,L)}
={\Tr\Bigl[ e^{- H/T}
\tilde n_{x,\alpha}(t) \tilde n_{0,\beta}(0)\Bigr]\over
\Tr\Bigl[e^{- H/T}\Bigr]}.
}
\end{eqnarray}
Here the trace $\Tr$ (which is different from $\Spur$ in (\ref{cc}) and
(\ref{nn})), is to be taken in the space ${\cal H}$, i.e.,
by definition,
\begin{equation} 
\Tr\Bigl[\cdots\Bigr]
=\sum_{\rm states}^{}
\frac{1}{L^N N^M}
\langle\Psi_{N,M}(k;\la)|\cdots|\Psi_{N,M}(k;\la)\rangle.
\end{equation}
The tildes over the Hubbard operators in the numerators in (\ref{ccnn})
mean the contraction of these operators onto the space ${\cal H}$,
i.e.,
\begin{equation} \label{PcP}
\tilde c^{}_{x,\alpha}\equiv {\cal P} c^{}_{x,\alpha} {\cal P},\qquad
\tilde c^\t_{x,\alpha}\equiv {\cal P} c^\t_{x,\alpha} {\cal P},\qquad
\tilde n^{}_{x,\alpha}\equiv {\cal P} n^{}_{x,\alpha} {\cal P}.
\end{equation}
The time dependence is described by the relations similar to
(\ref{time}) with the effective hamiltonian $H$ given by (\ref{Heff}).
The equations (\ref{ccnn}) exhibit the recipe for calculation of the
correlators in the strong coupling limit. Indeed, all states belonging
to the subspace ${\cal F}\backslash {\cal H}$ (of dimention $4^L-3^L$)
in the decomposition of the unit operator in the Fock space ${\cal F}$
inserted between canonical field operators in the numerators in
(\ref{cc}) do not contribute to the correlators due to infinitely
strong oscilation for any small (but finite) value of $t$ as $U$ tends
to infinity. The same happens with the correlators of density
operators, but since $[{\cal P}, n_{x,\alpha}(t)]=0$ such intermediate
states do not contribute already due to the temperature exponential.
Thus, equations (\ref{ccnn}) together with (\ref{PcP}) and
(\ref{P}) allow one to express the correlators in the strong coupling
limit through matrix elements (form factors) of Hubbard operators
$c^{}_{x,\alpha}(t)$,  $c^\t_{x,\alpha}(t)$, $n^{}_{x,\alpha}(t)$
between two states from ${\cal H}$ only.
It is to be emphasized that equations (\ref{ccnn}) will produce
the correct answers also for the equal-time correlation functions
at $U=\infty$ (since $t$ should be put equal to zero after taking the
limit $U\to \infty$).

The equal-time correlators of the canonical field operators
(given by the first and the second equations in (\ref{ccnn}) at $t=0$)
should satisfy simple relations governed by the anticommutation
relations between $\tilde c^{}_{x,\alpha}$ and $\tilde c^\t_{y,\beta}$.
If $x\ne y$ then $\tilde c^{}_{x,\alpha}$ and $\tilde c^\t_{y,\beta}$
anticommute, while at $x=y$ and $\alpha=\beta$ the anticommutators
are different from (\ref{anti}) being equal to
\begin{equation} 
[\tilde c^{}_{x,\up},\tilde c^\t_{x,\up}]_{+}
={\cal P}-\tilde n^{}_{x,\down},\qquad
[\tilde c^{}_{x,\down},\tilde c^\t_{x,\down}]_{+}
={\cal P}-\tilde n^{}_{x,\up}.
\end{equation}
In particular, it follows from these relations that
\begin{equation} \label{ccPnn}
\tilde c^{}_{x,\up}\tilde c^\t_{x,\up}
={\cal P}-(\tilde n^{}_{x,\up}+\tilde n^{}_{x,\down})
\end{equation}
(and the same for $\tilde c^{}_{x,\down}\tilde c^\t_{x,\down}$).
Having in mind that
$\tilde c^\t_{x,\alpha}\tilde c^{}_{x,\alpha}=\tilde n^{}_{x,\alpha}$,
and that ${\cal P}$ play the role of the unit operator in the space
${\cal H}$ it is easy to see what the relations between
correlators at $t=0$, $x=0$ are. Let us note that these relations are
different from those which one gets from (\ref{cc}) and (\ref{anti}).
It means that the limits $U\to \infty$ and $t\to 0$ do not ``commute''.

The operators $\tilde c^{}_{x,\alpha}$ and $\tilde c^\t_{x,\alpha}$, due to
(\ref{P}), can be also defined by means of their decompositions,
e.g., one has
\begin{eqnarray} \label{tilde_c}
\lefteqn{
\tilde c^{}_{x,\up} =
\sum_{N=0}^{L-1} \sum_{M=0}^{N}
\sum_{q_1<\dots<q_N\atop \mu_1<\dots<\mu_M}^{}
\sum_{k_1<\dots<k_{N+1}\atop \la_1<\dots<\la_M}^{}
}
\nonumber\\ &&
|\Psi_{N,M}(q;\mu)\rangle
\frac{{\cal F}^{(N+1,M)}_{x,\up}(q;\mu |k;\la)}{L^{2N+1} N^M (N+1)^M}
\langle\Psi_{N+1,M}(k;\la)|,
\end{eqnarray}
where ${\cal F}^{(N+1,M)}_{x,\up}(q;\mu |k;\la)$ are matrix elements
(form factors) of the operator $c^{}_{x,\up}$ between two states from
${\cal H}$:
\begin{equation} 
{\cal F}^{(N+1,M)}_{x,\up}(q;\mu |k;\la)\equiv
\langle\Psi_{N,M}(q;\mu)|c^{}_{x,\up}|\Psi_{N+1,M}(k;\la)\rangle.
\end{equation}
Explicit calculation results in the following representation
\begin{eqnarray} \label{ff}
\lefteqn{
{\cal F}^{(N+1,M)}_{x,\up}(q;\mu |k;\la)
=e^{-i\left({\Lambda-\Theta\over 2}\right)N}\,
{\det}_{M} W \, {\det}_{N+1} D
}
\nonumber\\ &&
\exp{\left\{ix\sum_{a=1}^{N+1}k_a-ix\sum_{b=1}^{N}q_b\right\}},
\end{eqnarray}
where $\Lambda=\sum_{a=1}^{M}\la_a$ and $\Theta=\sum_{b=1}^{M}\mu_b$.
Elements of the $(N+1)\times (N+1)$ matrix $D$ are ($a=1,\dots,N+1$;
$b=1,\dots,N$)
\begin{equation} 
(D)_{ab}=\sin\Bigl(\frac{\Lambda-\Theta}{2}\Bigr)
\cot\Bigl(\frac{k_a-q_b}{2}\Bigr),\qquad
(D)_{a,N+1}=1,
\end{equation}
and elements of the $M\times M$ matrix $W$ are ($a,b=1,\dots,M$)
\begin{equation} 
(W)_{ab}=\sum_{n=1}^{N} e^{in(\la_a-\mu_b)}.
\end{equation}
The similar representation can be obtained for the form factors of operator
$c^{}_{x,\down}$. The form factors of operators $c^\t_{x,\alpha}$ can be
obtained from the form factors of operators $c^{}_{x,\alpha}$ by means
of complex conjugation. The form factors of density operators
$n_{x,\alpha}$ can be also calculated; the corresponding results are
given in our recent paper \cite{AIP-97}.

Correlation functions (\ref{ccnn}) can be expressed, due to (\ref{tilde_c}),
as the sums over all intermediate states of squared modula of the
corresponding form factors. The resulting expressions for the
correlators are quite similar to those obtained earlier for the
two-component impenetrable gas \cite{IP-NPB}. On a finite lattice, they
are rather bulky, being considerably simplified in the thermodynamic
limit.

\section{Correlation functions in the thermodynamic limit}

In this Section the results of our calculation of the two-point
temperature correlation functions in the Hubbard model at $U=\infty$ in
the thermodynamic limit, $L\to\infty$, are given. To derive the
representations below from the representations on a finite
lattice we use the technique described in detail (for the two-component
gas) in paper \cite{IP-NPB}. In the limit,
the sums over the intermediate states in the expressions for the
correlators on a finite lattice can be reduced to expansions of the
Fredholm determinants of some linear integral operators.

Thus, the correlation functions considered are
represented as Fredholm determinants of linear integral operators.
The integral operators which enter the representations for the correlators
act on arbitrary function $f(k)$ according to the rule:
\begin{equation} 
(\hat {\cal A}\cdot f)(k)
= \int\limits_{-\pi}^{\pi} dk'\, {\cal A}(k,k') f(k'),
\end{equation}
where the function ${\cal A}(k,k')$ is the kernel of the integral operator
$\hat {\cal A}$.

In order to write down the representations for the correlators, let us
introduce some notations. Define the functions
\begin{eqnarray} \label{EG}
\lefteqn{
E(k)=E(k;x,t)
=\frac{1}{2\pi}\int\limits_{-\pi}^{\pi} dq
\frac{e^{-it\eps(q)+ixq}-e^{-it\eps(k)+ixk}}{\tan({q-k\over 2})},
}
\nonumber\\
\lefteqn{
G=G(x,t)=
\frac{1}{2\pi}\int\limits_{-\pi}^{\pi} dq e^{-it\eps(q)+ixq}.
}
\end{eqnarray}
The distance $x$ is an arbitrary integer; the dispersion is
$\eps(q)=-2\cos q$ (see (\ref{eeps})) where the momenta $q$
and $k$ in (\ref{EG}) take any values on
the interval $[-\pi,\pi]$.
In particular, $G(x,t)=i^x J_{x}(2t)$ where $J_{x}$ is the
Bessel function ($G(x,0)=\delta_{x,0}$).

The following pair of functions play an important role:
\begin{eqnarray} 
\lefteqn{
\ell_{+}(\eta|k)
=\left\{\frac{1-\cos\eta}{2}\, E_{+}(k) +
\frac{\sin\eta}{2}\, \frac{1}{E_{-}(k)}
\right\} \sqrt{\vartheta(k)},
}
\nonumber\\
\lefteqn{
\ell_{-}(k)
=E_{-}(k)\sqrt{\vartheta(k)},
}
\end{eqnarray}
where $\eta\in [-\pi,\pi]$ is a parameter and $\vartheta (k)$ is the Fermi
weight,
\begin{equation} 
\vartheta (k)=\frac{e^{-B/T}}{2\cosh{B\over T}+e^{(\eps(k)-h)/T}}.
\end{equation}
Here $h$ is the chemical potential and $B$ is the external constant
field. The functions $E_{\pm}(k)$ are
\begin{equation} 
E_{+}(k)=E(k)\, E_{-}(k),\qquad
E_{-}(k)=\exp\left(\frac{it\eps(k)-ixk}{2}\right).
\end{equation}
We introduce also the function
\begin{equation} 
F(\gamma;\eta):=
1+ \sum_{p=1}^{\infty} \gamma^{-p} (e^{ip\eta}+e^{-ip\eta}),
\end{equation}
where
\begin{equation} 
\gamma=1+e^{2B/T}.
\end{equation}
It is worth mentioning that $\gamma\in [1,\infty)$ for any real external
field $B$. At the point $\gamma=1$ ($B=-\infty$)
one has $F(1;\eta)=2\pi\Delta(\eta)$ where $\Delta(\eta)$
is the $2\pi$-periodic delta-function.

Now we are ready to formulate the main results.

For the correlation function of the canonical field operators of the
Hubbard model at $U=\infty$ we obtain the following representations:
\begin{eqnarray} \label{cc-dets}
\lefteqn{
\langle c^\t_{x,\up}(t) c^{}_{0,\up}(0) \rangle^{(T)}
=e^{-it(h-B)} \frac{1}{2\pi}\int\limits_{-\pi}^{\pi}d\eta\, F(\gamma;\eta)
}
\nonumber\\&&
\biggl[
\det\left(\hat{\cal I}+\gamma\,\hat{\cal Q}(\eta)+\hat{\cal R}^{(-)}\right)
-\det\left(\hat{\cal I}+\gamma\,\hat{\cal Q}(\eta)\right)
\biggr],
\nonumber\\
\lefteqn{
\langle c^{}_{x,\up}(t) c^\t_{0,\up}(0) \rangle^{(T)}
=e^{it(h-B)} \frac{1}{2\pi}\int\limits_{-\pi}^{\pi}d\eta\, F(\gamma;\eta)
}
\nonumber\\&&
\biggl[
\det\left(\hat{\cal I}+\gamma\,\hat{\cal Q}(\eta)
-\gamma\,\hat{\cal R}^{(+)}(\eta)\right)
+(G-1)\det\left(\hat{\cal I}+\gamma\,\hat{\cal Q}(\eta)\right)
\biggr],
\end{eqnarray}
where $\hat{\cal I}$ is the unit operator on the interval $[-\pi,\pi]$,
and
\begin{equation} \label{Q}
\hat{\cal Q}(\eta)\equiv
\hat{\cal V}(\eta)-\frac{1-\cos\eta}{2}\, G\, \hat{\cal R}^{(-)}.
\end{equation}
The integral operators $\hat{\cal V}(\eta)$, $\hat{\cal R}^{(-)}$,
$\hat{\cal R}^{(+)}(\eta)$ (the latter two are of rank one)
possess kernels
\begin{eqnarray} 
\lefteqn{
{\cal V}(\eta|k,k')
=\frac{\ell_{+}(\eta|k)\,\ell_{-}(k')-\ell_{-}(k)\,\ell_{+}(k')}
{2\pi\tan({k-k'\over 2})},
}
\nonumber\\
\lefteqn{
{\cal R}^{(-)}(k,k')=
\frac{1}{2\pi}\,\ell_{-}(k)\,\ell_{-}(k'),
}
\nonumber\\
\lefteqn{
{\cal R}^{(+)}(\eta|k,k')=
\frac{1}{\pi(1-\cos\eta)}\,\ell_{+}(\eta|k)\,\ell_{+}(\eta|k').
}
\end{eqnarray}
The representations for the correlators
$\langle c^\t_{x,\down}(t) c^{}_{0,\down}(0) \rangle^{(T)}$ and
$\langle c^{}_{x,\down}(t) c^\t_{0,\down}(0) \rangle^{(T)}$ can be
obtained from (\ref{cc-dets}) by inversing the sign of the
external field, $B\to -B$.

In the equal-time case, $t=0$, the integrals over $\eta$ in (\ref{cc-dets})
can be taken explicitly. For $t=0$ and $x\ne 0$ one has
\begin{eqnarray} 
\lefteqn{
\langle c^\t_{x,\up} c^{}_{0,\up}\rangle^{(T)} =
-\langle c^{}_{-x,\up} c^\t_{0,\up}\rangle^{(T)}=
}
\nonumber\\&&
=\det\left(\hat{\cal I}+ (\gamma-1) \hat v +\hat r\right)
-\det\left(\hat{\cal I}+ (\gamma-1) \hat v \right),
\end{eqnarray}
where the kernels of integral operators $\hat v$ and $\hat r$ are
\begin{eqnarray} 
\lefteqn{
v(k,k')=-\sqrt{\vartheta(k)}
\frac{\sin(|x|{k-k'\over 2})}{2\pi\tan({k-k'\over 2})}
\sqrt{\vartheta(k')},
}
\nonumber\\
\lefteqn{
r(k,k')=\sqrt{\vartheta(k)}
e^{-ix{k+k'\over 2}} \sqrt{\vartheta(k')}.
}
\end{eqnarray}
For $t=0$ and $x=0$ one has
\begin{eqnarray} 
\lefteqn{
\langle c^\t_{x,\up} c^{}_{x,\up}\rangle^{(T)}
=\frac{1}{2\pi}\int\limits_{-\pi}^{\pi}dk\, \vartheta(k),
}
\nonumber\\
\lefteqn{
\langle c^{}_{x,\up} c^\t_{x,\up}\rangle^{(T)}
=1-\frac{\gamma}{2\pi} \int\limits_{-\pi}^{\pi}dk\, \vartheta(k),
}
\end{eqnarray}
and therefore, e.g.,
\begin{equation} 
\langle c^{}_{x,\up} c^\t_{x,\up}\rangle^{(T)}=
1-\left(\langle c^\t_{x,\up} c^{}_{x,\up}\rangle^{(T)}
+\langle c^\t_{x,\down} c^{}_{x,\down}\rangle^{(T)}
\right),
\end{equation}
in agreement with the relation (\ref{ccPnn}).

Consider now the temperature
correlation functions of the density operators.
For these correlation functions the following representation is valid
\begin{eqnarray} \label{nn-dets}
\lefteqn{
\langle n_{x,\alpha}(t) n_{0,\beta}(0) \rangle^{(T)}
=\frac{1}{2\pi}\int\limits_{-\pi}^{\pi} d\eta\,
\Phi^{(\alpha,\beta)}(\gamma;\eta)
}
\nonumber\\&&
\frac{2}{1-\cos\eta}
\biggl[\det\left(\hat{\cal I}+\gamma\,\hat{\cal U}(\eta)\right)
-\det\left(\hat{\cal I}+\gamma\,\hat{\cal Q}(\eta)\right)\biggr],
\end{eqnarray}
where
\begin{eqnarray} 
\lefteqn{
\Phi^{(\up,\up)}(\gamma;\eta)
=\frac{\gamma-1}{\gamma^2}+2\pi\Delta(\eta)\,\frac{1}{\gamma^2},
}
\nonumber\\
\lefteqn{
\Phi^{(\up,\down)}(\gamma;\eta)=
\Phi^{(\down,\up)}(\gamma;\eta)
=-\frac{\gamma-1}{\gamma^2}+2\pi\Delta(\eta)\,\frac{\gamma-1}{\gamma^2},
}
\nonumber\\
\lefteqn{
\Phi^{(\down,\down)}(\gamma;\eta)=\frac{\gamma-1}{\gamma^2}
+2\pi\Delta(\eta)\,\left(\frac{\gamma-1}{\gamma}\right)^2,
}
\end{eqnarray}
and $\Delta(\eta)$ is the $2\pi$-periodic delta-function. The integral
operator $\hat{\cal U}(\eta)$ possesses the kernel
\begin{equation} 
{\cal U}(\eta|k,k')
=\frac{\ell_{+}(\eta|k)\,\ell_{-}(k')-\ell_{-}(k)\,\ell_{+}(k')}
{2\pi\sin({k-k'\over 2})},
\end{equation}
and the integral operator $\hat{\cal Q}(\eta)$ is defined in (\ref{Q}).
Let us note that the contributions to the correlators (\ref{nn-dets})
containing the delta-function $\Delta(\eta)$ in the quantities
$\Phi^{(\alpha,\beta)}(\gamma;\eta)$ admit more explicit form.
Indeed, only the Fredholm minors up to the second order
contribute to the correlators in these terms.
Let us denote this contribution (up to
a numerical factor depending on spins) as ${\cal G}(x,t;h,B)$. One has
\begin{eqnarray} 
\lefteqn{
{\cal G}(x,t;h,B):=
\frac{2}{1-\cos\eta}
\biggl[\det\left(\hat{\cal I}+\gamma\,\hat{\cal U}(\eta)\right)
-\det\left(\hat{\cal I}+\gamma\,\hat{\cal Q}(\eta)\right)\biggr]
\biggr|_{\eta=0}
}
\nonumber\\&&
=\left(\frac{\gamma}{2\pi}
\int\limits_{-\pi}^{\pi}dk\,\vartheta(k)\right)^2
-\left|\frac{\gamma}{2\pi}
\int\limits_{-\pi}^{\pi}dk\,\vartheta(k)\, e^{it\eps(k)-ixk}\right|^2
\nonumber\\&&
+\left(\frac{1}{2\pi}
\int\limits_{-\pi}^{\pi}dk\, e^{-it\eps(k)+ixk}\right)
\left(\frac{\gamma}{2\pi}
\int\limits_{-\pi}^{\pi}dk\,\vartheta(k)\, e^{it\eps(k)-ixk}\right).
\end{eqnarray}
Other contributions to the correlators (terms independent on $\eta$
in $\Phi^{(\alpha,\beta)}(\gamma;\eta)$)
could not be simplified in the general
case since the integral over $\eta$ cannot be taken in the closed form.
One can, however, consider some combinations of the correlators
(\ref{nn-dets}) in which these terms cancel; e.g.,
\begin{equation} 
\langle \left(n_{x,\up}(t)+n_{x,\down}(t)\right)
\left(n_{0,\up}(0)+n_{0,\down}(0)\right)\rangle^{(T)}
={\cal G}(x,t;h,B).
\end{equation}

In the equal-time case
the integral over $\eta$ in (\ref{nn-dets}) can be taken for all the
contributions. Remarkably that in the case $t=0$ and $x\ne 0$
the correlation functions are determined by the function
${\cal G}(x,t;h,B)$ only. In this case one has
\begin{equation} 
\langle n_{x,\alpha} n_{0,\beta} \rangle^{(T)}
= \phi^{(\alpha,\beta)}
\Biggl\{
\left(\frac{\gamma}{2\pi}
\int\limits_{-\pi}^{\pi}dk\,\vartheta(k)\right)^2
-\left|\frac{\gamma}{2\pi}
\int\limits_{-\pi}^{\pi}dk\,\vartheta(k)\, e^{-ixk}\right|^2
\Biggr\}
\end{equation}
where
\begin{equation} 
\phi^{(\up\up)}=\frac{1}{\gamma^2},\qquad
\phi^{(\up\down)}=\phi^{(\down\up)}=\frac{\gamma-1}{\gamma^2},\qquad
\phi^{(\down\down)}=\left(\frac{\gamma-1}{\gamma}\right)^2.
\end{equation}
In the case $t=0$ and $x=0$ not only the function ${\cal G}(x,t;h,B)$
contributes to the correlators. The result is
\begin{eqnarray} 
\lefteqn{
\langle n_{x,\up} n_{x,\up} \rangle^{(T)}
=\langle n_{x,\up} \rangle^{(T)}
=\frac{1}{2\pi} \int\limits_{-\pi}^{\pi}dk \,\vartheta(k),
}
\nonumber\\
\lefteqn{
\langle n_{x,\down} n_{x,\down} \rangle^{(T)}
=\langle n_{x,\down} \rangle^{(T)}
=\frac{\gamma-1}{2\pi} \int\limits_{-\pi}^{\pi}dk \,\vartheta(k),
}
\nonumber\\
\lefteqn{
\langle n_{x,\up} n_{x,\down} \rangle^{(T)}
=\langle n_{x,\down} n_{x,\up} \rangle^{(T)}=0.
}
\end{eqnarray}

Let us note that the obtained results for the correlation functions
(\ref{cc-dets}) and (\ref{nn-dets}) has the proper ``one-component
limit''.  Indeed, in the limit $B\to -\infty$, $h\to -\infty$
($h-B=h_0$ is fixed) one has a free fermion model (of fermions with spin
$\up$).  Since $\gamma=1$ in this limit, the representations
(\ref{cc-dets}) and (\ref{nn-dets}) become in fact trivial reproducing
the well-known results for the correlators of free fermions on the
lattice.  Also for the only non-vanishing correlator of density
operators, $\langle n_{x,\up}(t) n_{0,\up}(0)\rangle^{(T)}$,
one gets in this limit the expression for the correlator
of third local spin components in the $XX0$ spin chain
\cite{N-67,CIKT-93}.

In conclusion, we would like to stress that the integral operators
involved into the representations (\ref{cc-dets}) and (\ref{nn-dets})
are of the form usual for integrable models, i.e.,
they are ``integrable integral operators'' \cite{IIKS-93,KBI}.
This fact is important for constructing the corresponding
matrix Riemann-Hilbert problem and for deriving integrable partial
differential equations for the correlators.  This, in turn,
will make possible the evaluation of different (e.g., large time and
distance) asymptotics of the correlators considered.

\section*{Acknowlegments}

This work is supported in part by the grant INTAS-RFBR 95-0414.

\end{document}